\documentclass[aps,prc,fleqn,floatfix,twocolumn,superscriptaddress,twoside]{revtex4}

\usepackage[dvipdfm]{graphicx}

\usepackage{amsmath,amssymb,amsfonts}

\usepackage{color}

\begin{document}

\title{Jet energy loss, photon production, and photon-hadron correlations at RHIC}
\author{Guang-You Qin}
\affiliation{Department of Physics, The Ohio State University, Columbus, OH, 43210, USA}

\author{J\"org Ruppert}
\author{Charles Gale}
\author{Sangyong Jeon}

\affiliation{Department of Physics, McGill University, Montreal, Quebec, H3A 2T8, Canada}

\author{Guy D. Moore}

\affiliation{Department of Physics, McGill University, Montreal, Quebec, H3A 2T8, Canada}

\affiliation{Universidad Autonoma de Madrid, E-28049 Madrid, Spain}

\date{\today}
\begin{abstract}

Jet energy loss, photon production and photon-hadron correlations are
studied together at high transverse momentum in relativistic heavy-ion
collisions at RHIC energies. The modification of hard jets traversing a
hot and dense nuclear medium is evaluated by consistently taking into
account induced gluon radiation and elastic collisions. The production
of high transverse momentum photons in Au+Au collisions at RHIC is
calculated by incorporating a complete set of photon-production
channels. Comparison with  experimental photon production and
photon-hadron correlation data is performed, using a (3+1)-dimensional relativistic hydrodynamic description of the thermalized medium created in these collisions. Our results demonstrate that the interaction between the hard jets and the soft medium is important for the study of photon production and of photon-hadron correlation at RHIC.

\end{abstract}
\maketitle

\section{Introduction}

Hard processes are regarded as good tools for the tomographic study of relativistic heavy-ion collisions and of the quark-gluon plasma: The energetic probes are created in the early stage of the collisions and can therefore access the space-time history of the transient
hot and dense nuclear medium created in these collisions. Large transverse momentum partons have been identified as especially useful because they interact directly with the nuclear medium, and will lose energy in the process. The partonic jets which emerge  will fragment into the hadrons which are observed in the
detectors. Experiments at the Relativistic Heavy Ion Collider (RHIC) at Brookhaven National Laboratory (BNL)  have indeed shown that high-$p_T$ hadrons in
central A+A collisions are significantly suppressed in comparison with those in binary-scaled p+p interactions  \cite{Adcox:2001jp, Adler:2002xw, Gyulassy:1993hr}. Those results are commonly referred to as ``jet-quenching''.

A lot of effort has been devoted to describe the energy loss experienced
by the color charges inside the hot, strongly interacting matter. Gluon
bremsstrahlung with the Landau-Pomeranchuk-Migdal (LPM) \cite{Migdal:1956tc} effect has been proposed as the dominant
mechanism for the jet-quenching process, and different theoretical formalisms have been elaborated to describe the hard
parton energy loss \cite{Baier:1996kr, Kovner:2003zj, Zakharov:1996fv, Gyulassy:2000er, Wang:2001ifa, Zhang:2003yn, Majumder:2004pt, Majumder:2007hx, Arnold:2001ms, Arnold:2001ba, Arnold:2002ja}. Several of these bremsstrahlung calculations were compared against each other \cite{Qin:2007zz, Renk:2006sx, Majumder:2007ae, Bass:2008rv}, using a relativistic ideal (3+1)-dimensional hydrodynamical simulation of the strongly interacting medium \cite{Nonaka:2006yn}. This implementation improved upon earlier attempts relying on simple schematic models by building on a detailed and realistic description of the time evolution. Another potentially important energy loss mechanism for high-$p_T$ color charges is provided by scattering off thermal
partons, in binary elastic collisions. Estimations of the relative magnitude of collisional versus radiative energy loss have also been performed in several different approaches and scenarios \cite{Bjorken:1982tu, Mustafa:2003vh, Mustafa:2004dr, Adil:2006ei, Wicks:2007zz, Zakharov:2007pj, Renk:2007id, Majumder:2008zg}. In Ref. \cite{Qin:2007rn}, a  study of radiative and collisional energy loss was carried out in a given single approach (AMY \cite{Arnold:2001ms, Arnold:2001ba, Arnold:2002ja}). Note that the collisional energy loss was further improved upon in Ref. \cite{Schenke:2009ik}.

Complementary to hadronic species, photons represent another set of
promising observables with the potential of probing the quark-gluon
plasma (see, for example, \cite{Gale:2009gc} and references therein). Since they carry no color charge, once produced, they will
essentially escape from the collision zone without further interaction
with the surrounding nuclear medium, reflecting directly local medium
properties. However, the experimental observation of produced photons in
high energy nuclear collisions involves the entire evolution of the
collision, together with a variety of sources. Considering that the
background contribution from meson decay may in principle be
experimentally subtracted, theoretical studies need to include photons
produced in the early collisions, those involving jet-medium
interactions during jet propagation in the medium \cite{Fries:2002kt}, and those from fragmentation of the surviving jets after their escape from the medium. In Ref. \cite{Turbide:2005fk, Turbide:2007mi}, the calculation of photon production from nuclear collisions at RHIC with those different photon sources 
has been done by first employing a (1+1)-dimensional Bjorken evolution model, and then  (2+1)-dimensional relativistic
hydrodynamics.

These calculations, and others, have given rise to the hope that the ``tomography'' of the hot and dense core of the nuclear medium created in relativistic heavy-ion collisions may be put on a firm quantitative footing.

In what concerns hadrons, several quantities have been put forward as ``tomographic variables''. One of those is the nuclear modification
factor $R_{AA}$ which is inferred from the measurement of single-particle inclusive $p_T$ spectra. It is
defined as the ratio of the hadron yield in A+A collisions to that in
binary-scaled p+p interactions. Since the single-particle spectrum is a
convolution of the jet production cross section and jet fragmentation
functions, the suppression of the produced hadrons at a fixed $p_T$
involves jets with a wide range of initial transverse energies; for
details see \cite{Qin:2007zz, Qin:2007rn}. In retrospect, it is
therefore not particularly surprising that a wide variety of energy loss
conjectures that differ significantly in the predicted energy loss
mechanism can reproduce the measured $R_{AA}$ in central Au+Au collisions at $\sqrt{s}=200~{\rm AGeV}$ at RHIC given the present experimental errors for central collisions. In that sense, the tomographic usefulness of $R_{AA}$ in central collisions has become rather questionable \cite{Renk:2006qg}. There have been several suggestions to improve the situation. One is to consider a more differential observable, and study $R_{AA}$ in non-central collisions as a
function of the azimuth and $p_T$ \cite{Majumder:2006we}. This effectively corresponds to studying an average over different paths of partons as they traverse the expanding medium, which is no longer rotationally invariant in the transverse plane. In Ref. \cite{Qin:2007zz}, $R_{AA}$ was also studied as a function of $p_T$ in central and
non-central collisions at mid and forward rapidities. Although this does
not restrict the initial jets' energies or
initial vertices, it allows for a different averaging over the medium than in central collisions and can thus
provide us with additional information.

Another road to improvement consists of considering many-body variables, e.g. to study the production of
high-$p_T$ hadrons correlated with other high-$p_T$ particles. One motivation
is that the correlation between the trigger particle and the measured hadrons will constrain the initial energy
distribution of the parton that fragments into the hadrons more tightly than single-particle measurements. 
One such suggested trigger is a hadron, see e.g.,
\cite{Renk:2006pk}. Note however that choosing a specific
$p_T$ for the trigger hadron on the ``near-side'' still does not
strictly pin down the energy of the ``away-side'' partons. The trigger
hadron is produced by fragmentation of partons that have also traversed
the medium and lost energy. Therefore, a specific trigger bias is
introduced into the away-side parton distribution and initial vertex
distribution. This effect has been explored in \cite{Renk:2006pk}, where
a detailed analysis of the initial vertex density and conditional
probability distribution $P(p_T)$ of away-side partons for di-hadron correlations was presented in the Baier-Dokshitzer-Mueller-Peigne-Schiff (BDMPS) framework of energy loss \cite{Baier:1996kr} by  gluon bremsstrahlung. There it was shown that the yield of hadrons per trigger hadron was sensitive to the specific assumptions about the evolution model for di-hadron correlations, while it was not for $R_{AA}$.

Another promising trigger is a high-$p_T$ photon: One studies jet-quenching by measuring the $p_T$ distribution of charged hadrons in the opposite direction of a trigger direct photon \cite{Wang:1996yh,Wang:1996pe}.
Direct photons are predominantly produced from hard collisions in the early stage of relativistic nuclear collisions.
Assuming that a high-$p_T$ direct photon from these processes can be used as a trigger of the away-side hadron, the transverse energy of the initial away-side parton before energy loss can then be deduced (up to corrections at next to leading order in $\alpha_{s}$). Thus, direct photons can provide a calibrated probe for the study of the properties of high energy density QCD.
This kind of trigger does not introduce a vertex bias and therefore
weighs all vertices according to the nuclear overlap function; and it strongly restricts the initial away-side partons' energy to a mono-energetic source. Given this restriction of the initial jet's transverse energy, the expected tomographic capabilities of photon-tagged jet correlation measurements are much higher than those of $R_{AA}$. It has been shown in Ref. \cite{Renk:2006qg} that while different schematic conjectures of jet energy losses could not be discriminated by $R_{AA}$ in central collisions at mid-rapidity, they gave clearly different predictions for $\gamma$-hadron correlations. Some other studies have also been performed along this direction \cite{Filimonov:2005kp, Arleo:2007qw, Zhang:2009rn}.

However, jet-photon correlation calculations need to go beyond direct photons and to include the other important and known electromagnetic sources such as fragmentation and jet-medium interaction. As the initial away-side partonic jets in these processes could have larger transverse energies than the near-side trigger photon, photons produced from those processes might have significant contribution to the correlations between final photons and hadrons. In this work, we consider these possibilities and incorporate a complete set of high-$p_T$ photon-production channels and study their relative contributions to the correlations between back-to-back photons and hadrons.

In Ref. \cite{Qin:2007rn}, we have presented a consistent calculation of both collisional and radiative energy loss in the same formalism. There it was applied to calculate the nuclear modification factor $R_{AA}$ of neutral pions in heavy ion collisions at RHIC for different centralities at mid-rapidity. Here, we continue and extend this effort by studying the effect of jet-medium interaction on the photon production as well as on correlated back-to-back hard photons and hadrons in high energy nuclear collisions. We employ the formalism developed in Ref. \cite{Qin:2007rn} to account consistently for collisional and radiative energy loss of the leading hard parton while it traverses the surrounding soft nuclear matter, modeled by (3+1)-dimensional hydrodynamics \cite{Nonaka:2006yn}. 

This paper is organized as follows: the next two sections describe our approach to jet energy loss and to photon production. Then, photon-hadron correlations are treated, and comparisons with experimental results obtained at RHIC by the PHENIX and STAR collaborations are performed. 

\section{Jet energy loss}

In our approach, hard jets (quarks and gluons) evolve in the soft nuclear medium according to a set of Fokker--Planck
type rate equations for their momentum distributions $P(E,t) = {dN(E,t)}/{dE}$. The generic
form of these rate equations can be written as follows \cite{Jeon:2003gi,Turbide:2005fk},
\begin{eqnarray}\label{FP-eq}
\frac{dP_j(E,t)}{dt} \!&=&\! \! \sum_{ab} \! \int \! d\omega \left[P_a(E+\omega,t) \frac{d\Gamma_{a\to
j}(E+\omega,\omega)}{d\omega dt} \right. \nonumber\\ && \left. - P_j(E,t)\frac{d\Gamma_{j\to b}(E,\omega)}{d\omega
dt}\right], \ \ \ \ \ \
\end{eqnarray}
where $d{\Gamma_{j\to a}(E,\omega)}/{d\omega dt}$ is the transition rate for the partonic process $j\to a$, with $E$
the initial jet energy and $\omega$ the energy lost in the process. The $\omega<0$ part of the integration
accounts for the contribution from the energy-gain channels. The radiative and collisional parts of the transition rates have been extensively discussed in Ref. \cite{Qin:2007zz,Qin:2007rn,Qin:2008ea}.

The initial jet momentum profiles may be computed from perturbative QCD calculations \cite{Owens:1986mp},
\begin{eqnarray}\label{initial_jet}
\frac{d\sigma_{AB\to jX}}{d^2p_T^j dy} &=& K_{\rm jet} \sum_{abd} \int dx_a G_{a/A}(x_a,Q)G_{b/B}(x_b,Q)\nonumber \\ && \times
\frac{1}{\pi} \frac{2x_ax_b}{2x_a-x_T^j e^y} \frac{d\sigma_{ab\to jd}}{dt}, \ \ \
\end{eqnarray}
with $x_T^j = 2p_T^j/\sqrt{s_{NN}}$, where $\sqrt{s_{NN}}$ is the center-of-mass energy. {In the above equation,
$G_{a/A}(x_a,Q)$ is the distribution function of parton $a$ with momentum fraction $x_a$ in the nucleus $A$ at
factorization scale $Q$, taken from CTEQ5 \cite{Lai:1999wy} including nuclear shadowing effects {from EKS98}
\cite{Eskola:1998df}. The distribution ${d\sigma}/{dt}$ is the leading order QCD differential cross section with $K_{\rm jet}$  accounting for next-to-leading order (NLO) corrections. It is fixed following  Ref. \cite{Eskola:2005ue} by reproducing the experimental measurement of the inclusive $\pi^0$ cross section at high-$p_T$ in in p+p collisions at $\sqrt{s_{NN}}=200~{\rm GeV}$ (see Fig.\ 1 in Ref.\cite{Qin:2007zz}). The $K_{\rm jet}$-factor is found to be $2.8$ when the renormalization scale and the factorization scale are taken to be the transverse momentum of the initial jets, and the fragmentation scale is taken to be the transverse momentum of produced hadrons.

To obtain the final high-$p_T$ hadron spectra in Au+Au collisions at RHIC, the energy loss of partonic jets in the nuclear medium must be taken into account. This is performed by calculating the medium-modified fragmentation function
$\tilde{D}_{h/j}(z,\vec{r}_\bot, \phi)$ for a single partonic jet,
\begin{eqnarray}
\label{mmff} \tilde{D}_{h/j}(z,\vec{r}_\bot, \phi) \!&=&\!\! \sum_{j'} \!\int\! dp_{j'} \frac{z'}{z}
D_{h/j'}(z') P(p_{j'}|p_j,\vec{r}_\bot, \phi), \ \ \ \
\end{eqnarray}
where $z = p_h / p_{j}$ and $z' = p_h / p_{j'}$ are two momentum fractions with $p_h$ the hadron momentum and $p_{j}$($p_{j'}$) the initial (final) jet momentum; $D_{h/j}(z)$ is the vacuum fragmentation function, taken from the KKP parametrization \cite{Kniehl:2000fe}. In the above equation, the probability function $P(p_{j'}|p_j,\vec{r}_\bot,
\phi)$ is obtained by solving Eq.~(\ref{FP-eq}), representing the probability of obtaining a jet $j'$ with momentum
$p_{j'}$ from a given jet $j$ with momentum $p_j$. This depends on the path taken by the parton and the medium
profile along that path, which in turn depends on the original location
of the jet, $\vec{r}_\bot$, and on its propagation angle
$\phi$ with respect to the reaction plane. Therefore, one must convolve the above expression over all transverse
positions and directions to obtain the final hadron spectra:
\begin{eqnarray}
\frac{d\sigma_{AB\to hX}}{d^2p_T^hdy} \!&=&\! \frac{1}{2\pi} \int d^2\vec{r}_\perp {\cal
P}_{AB}(b,\vec{r}_\perp) \nonumber \\ && \times \sum_{j} \int \frac{dz}{z^2} \tilde{D}_{h/j}(z,\vec{r}_\bot, \phi) \frac{d\sigma_{AB\to jX}}{d^2p_T^j dy}, \ \ \ \ \ \
\end{eqnarray}
where ${\cal P}_{AB}(b,\vec{r}_\perp)$ is the probability distribution of initial hard jets in the transverse position
$\vec{r}_\perp$, which is determined from the overlap geometry between two nuclei in the transverse plane of the collision zone. For A+B collisions at impact parameter $b$,
\begin{eqnarray}
\mathcal{P}_{AB}(b,\vec{r}_\bot) \!&=&\! \frac{T_A(\vec{r}_\bot + \vec{b}/2)T_B(\vec{r}_\bot - \vec{b}/2)}
{T_{AB}(b)}, \ \ \ \ \
\end{eqnarray}
where $T_A(\vec{r}_\bot)=\int dz \rho_A(\vec{r}_\bot,z)$ is the nuclear thickness function and $T_{AB}(b)=\int d^2r_\bot T_A(\vec{r}_\bot)
T_B(\vec{r}_\bot+\vec{b})$ is the nuclear overlap function. A Woods-Saxon form for the nuclear density function $\rho(\vec{r}_\bot,z)={\rho_0}/[{1+\exp(\frac{r-R}{d})}]$ is used, and the values of the parameters  $R=6.38~{\rm fm}$ and $d=0.535~{\rm fm}$ are taken from \cite{DeJager:1974dg}.

Putting all of the above together, one obtains the total yield of hadrons produced in relativistic nuclear collisions,
\begin{eqnarray}
\frac{dN^{h}_{AB}(b)}{d^2p_T^h dy} = \frac{N_{\rm coll}(b)}{\sigma_{NN}} \frac{d\sigma_{AB\to hX}}{d^2p_T^hdy}
\end{eqnarray}
where $N_{\rm coll}$ is the number of binary collisions and $\sigma_{NN}$ is the inelastic cross section of
elementary nucleon-nucleon collisions.
Finally, the nuclear modification factor $R_{AA}$ is defined as the ratio of the hadron yield in A+A collisions to that
in p+p interactions scaled by the number of binary collisions,
\begin{eqnarray}
R^h_{ AA}\! &=&\! \frac{1}{N_{\rm coll}(b)} \frac{{dN^h_{ AA}}(b)/{d^2p_T^hdy}} {{dN^h_{ pp}}/{d^2p_T^hdy}}.
\end{eqnarray}

\section{Photon production}

As mentioned already, there are several sources of high-$p_T$ photons  in relativistic nuclear
collisions. The most important are direct photons, fragmentation photons, and jet-medium photons (bremsstrahlung photons and conversion
photons).
The thermal photon emission arising from the partonic medium and later
hadronic phase makes a negligible contribution in the high-$p_T$ regime, and thus is excluded from consideration in the present calculation.

Direct photons are predominantly produced from early hard collisions between partons from the initial nuclei, via quark-anti-quark annihilation ($q+\bar{q} \to g+\gamma$) and quark-gluon Compton scattering ($q({\bar q})+g\to q({\bar q})+\gamma$). The inclusive direct photon cross section may be calculated from Eq.~(\ref{initial_jet}) by applying the corresponding partonic processes for photon production,
\begin{eqnarray}
\frac{d\sigma^{\rm direct}_{AB\to \gamma X}}{d^2p_T^\gamma dy} \!&=&\! K_{\gamma} \sum_{abd} \int dx_a G_{a/A}(x_a,Q)G_{b/B}(x_b,Q)
\nonumber \\ && \times \frac{1}{\pi} \frac{2x_ax_b}{2x_a-x_T^\gamma e^y} \frac{d\sigma_{ab\to \gamma d}}{dt}, \ \ \
\end{eqnarray}
with $x_T^\gamma = 2p_T^\gamma/\sqrt{s_{NN}}$.
The factor $K_{\gamma}$ is a function of the photon's transverse
momentum and is deduced  by performing an NLO calculation of photon
production in p+p collisions \cite{Aurenche:1987fs,Aversa:1988vb,
  Aurenche:1998gv}, in which the renormalization scale and the factorization scale are set to be the photon transverse momentum.

Fragmentation photons are produced by the surviving high energy jets after their passing through the hot and dense nuclear medium. The calculation of fragmentation photon spectra is similar to that of the high-$p_T$ hadron production described in the last section. We may also define a medium-modified photon fragmentation function $\tilde{D}_{\gamma/j}(z,\vec{r}_\bot, \phi)$ as in Eq.(\ref{mmff}). Then the final expression  for the fragmentation photon cross section can be written as
\begin{eqnarray}
\frac{d\sigma^{\rm frag}_{AB\to \gamma X}}{d^2p_T^\gamma dy} \!&=&\! \frac{1}{2\pi} \int d^2\vec{r}_\perp {\cal
P}_{AB}(\vec{r}_\perp) \nonumber\\ && \times \sum_{j} \int \frac{dz}{z^2} \tilde{D}_{\gamma/j}(z,\vec{r}_\bot, \phi) \frac{d\sigma_{AB\to jX}}{d^2p_T^j dy}. \ \ \ \ \ \
\end{eqnarray}
where the vacuum fragmentation functions of quarks and gluons into real photons are taken from \cite{Bourhis:1997yu}.

Jet-medium photons are produced by jet-medium interactions during the passage of jets through the hot nuclear medium. These include induced photon radiation (bremsstrahlung photons) and direct conversion from the high energy jets (conversion photons). It has been shown that those two processes are important for the understanding of experimental data for photon production in Au+Au collisions at RHIC
\cite{Turbide:2005fk, Turbide:2007mi, Fries:2002kt}.

In order to incorporate the photons directly produced from jet-medium
interaction, we may add to the set of evolution equations shown in Eq.\
(\ref{FP-eq}) an additional evolution equation for photons,
\begin{eqnarray}
\label{photon_evolve} \frac{dP^{\rm JM}_\gamma(E,t)}{dt} \!&=&\!\! \int \! d\omega
P_{q\bar{q}}(E{+}\omega,t) \frac{d\Gamma^{\rm JM}_{q\to \gamma}(E{+}\omega,\omega)}{d\omega dt}, \ \ \ \ \ \
\end{eqnarray}
where ${d\Gamma^{\rm JM}_{q\to \gamma}}/{d\omega dt} = {d\Gamma^{\rm brem}_{q\to \gamma}}/{d\omega dt} + {d\Gamma^{\rm conv}_{q\to \gamma}}/{d\omega dt}$. The transition rates $d\Gamma^{\rm brem}_{q\to \gamma}/d\omega dt$  for photon bremsstrahlung processes are similar to gluon bremsstrahlung processes and are extensively discussed in \cite{Arnold:2001ms, Arnold:2001ba, Arnold:2002ja}. And the transition rates $d\Gamma^{\rm conv}_{q\to \gamma}/d\omega dt$ for binary jet-photon conversion processes may be inferred from the photon emission rates for those processes \cite{Qin:2008rd},
\begin{eqnarray}
\frac{d\Gamma^{\rm conv}_{q\to \gamma}(E,\omega)}{d\omega dt} \!&=&\!\! \sum_f \left(\frac{e_f}{e}\right)^2 \frac{2\pi
\alpha_e \alpha_s T^2}{3E} \nonumber\\ && \times \left(\frac{1}{2} \ln \frac{ET}{m_q^2} + C_{2\to 2}\left({E}/{T}\right) \right)
\delta(\omega), \ \ \ \ \ \
\end{eqnarray}
where $m_q^2 = g_s^2 T^2 / 6$ is the thermal quark mass and $C_{2\to 2}
\approx -0.36149$ in the limit of $E \gg T$ \cite{Kapusta:1991qp, Baier:1991em, Arnold:2002ja}. 
The function $\delta(\omega)$ represents the fact that the produced photon has essentially the same energy as the incoming quarks (or anti-quarks) in the jet-photon conversion processes.


\section{Photon-hadron correlations}

In this section, we present the calculation of the correlation between back-to-back hard photons and hadrons.
The associated hadrons are produced from the fragmentation of surviving jets after their passing through the nuclear medium, while the trigger photons may come from various sources as has been discussed in the previous section. Therefore, the photon--hadron correlation will depend on the jet-photon correlation at the production vertex as well as on the energy loss of the jets during their propagation in the medium.

In correlation studies, one of the commonly exploited observables is the yield per
trigger, which is  the momentum distribution of  produced  hadrons on the away side,
given a trigger photon of momentum $p_T^\gamma$ in the near side. Following
Ref.~\cite{:2008cqb}, one may write 
\begin{eqnarray}
P(p_T^h|p_T^\gamma) = \frac{P(p_T^h,p_T^\gamma)}{P(p_T^\gamma)},
\end{eqnarray}
where $P(p_T^\gamma)=1/\sigma_{\rm tot}\int dy_\gamma d\sigma_\gamma / dy_\gamma
dp_T^\gamma$ represents the single-particle $p_T$ distribution and
$P(p_T^\gamma, p_T^{h})=1/\sigma_{\rm tot}\int dy_\gamma
dy_{h}d\sigma_{\gamma+h}/dy_\gamma dy_{h} dp_T^\gamma dp_T^{h}$ is the $\gamma-h$ pair
$p_T$ distribution.

The trigger photon and the associated hadron are produced from a pair of back-to-back partons. Assuming no correlation between the individual evolutions of two partonic jets once they are
produced, we may write down the expressions of these two distributions as follows
\begin{eqnarray} \label{1-particle}
P_f(p_T^\gamma) \!&=&\! \int \frac{d\phi}{2\pi}\! \int \! d^2\vec{r}_\bot {\cal P}_{AB}(\vec{r}_\bot) \nonumber \\&& \times \!\sum_{j} \! \int \! dp_T^{j} P_i(p_T^j)
P(p_T^\gamma|p_T^j,\vec{r}_\bot, \phi),  \ \ \ \ \ \
\nonumber \\
P_f(p_T^h,p_T^\gamma) \! &=&\! \int \frac{d\phi}{2\pi} \! \int \! d^2\vec{r}_\bot {\cal P}_{AB}(\vec{r}_\bot)
\nonumber \\ && \times
\!\sum_{jj'} \! \int \! dp_T^{j} dp_T^{j'} P_i(p_T^j, p_T^{j'})
\nonumber \\ && \times P(p_T^\gamma|p_T^j,\vec{r}_\bot, \phi) P(p_T^h|p_T^{j'},\vec{r}_\bot, \pi \!+\! \phi). \ \ \ \ \ \
\end{eqnarray}
In the above equations, $P_f(p_T^\gamma)$ and $P_f(p_T^h, p_T^{\gamma})$ are the
final state distribution functions for single-photon and back-to-back photon-hadron production,
while $P_i(p_T^j)$ and $P_i(p_T^j,p_T^{j'})$ are the initial distributions 
for a jet or jet pair to be produced in the medium with the given
momenta and species types.
The yield per trigger 
$P(p_T^\gamma|p_T^j,\vec{r}_\bot, \phi)$ is the number of produced photons 
given an initial jet, which may be decomposed into different contributions,
\begin{eqnarray}
P(p_T^\gamma|p_T^j,\vec{r}_\bot, \phi) = \sum_{\rm src} P(p_T^\gamma, {\rm src}|p_T^j,\vec{r}_\bot, \phi),
\end{eqnarray}
where the sources include direct photons, fragmentation photons and
jet-medium photons as described in the last section. 
For the direct photon contribution, the near-side jet is replaced by a direct photon, 
so the yield per trigger is simply given by 
$P(p_T^\gamma, {\rm direct}|p_T^j,\vec{r}_\bot,\phi)  = \delta(p_T^\gamma-p_T^j)$.
The fragmentation photon contribution is related to the medium-modified photon fragmentation function by
$P(p_T^\gamma, {\rm frag}|p_T^{j},\vec{r}_\bot, \phi) = \tilde{D}_{\gamma/j}(z,\vec{r}_\bot, \phi)/p_T^{j}$,
with $z=p_T^\gamma/p_T^{j}$.
Furthermore, we also need to include the contribution from jet-medium photons, which is achieved by solving Eq.~(\ref{photon_evolve}), the photon evolution equation, to obtain 
the yield per trigger $P(p_T^\gamma, {\rm JM}|p_T^j,\vec{r}_\bot,\phi)$. Note that for the case of
hadron yield per jet trigger $P(p_T^h|p_T^j,\vec{r}_\bot, \phi)$, we
only need to include the fragmentation contribution since we focus on high-$p_T$ hadron production.

For the study of photon-hadron correlations, it is often useful to define the photon-triggered fragmentation function as follows \cite{Wang:2003mm, Wang:2003aw},
\begin{eqnarray}
D_{AA}(z_T,p_T^\gamma) = p_T^\gamma P_{AA}(p_T^h|p_T^\gamma),
\end{eqnarray}
with $z_T = p_T^h/p_T^\gamma$. In order to quantify the effect of the nuclear medium on the photon-hadron correlations, we may also introduce the nuclear modification factor $I_{AA}$, which is defined as
the ratio between A+A and p+p collisions of the
photon-triggered fragmentation function,
\begin{eqnarray}
I_{AA}(z_T,p_T^\gamma) = \frac{D_{AA}(z_T,p_T^\gamma)}{D_{pp}(z_T,p_T^\gamma)}.
\end{eqnarray}

\section{Results at RHIC}

In the previous sections, we have outlined the theoretical formalism for
calculating jet energy loss, the production of high-$p_T$ hadrons and photons, and photon-hadron correlations.
In this section, we apply this methodology to study Au+Au collisions at
RHIC, using a (3+1)-dimensional relativistic ideal hydrodynamics
\cite{Nonaka:2006yn}, and compare our results to available experimental
measurements.

\begin{figure}[htb]
\begin{center}
\resizebox{1.0\linewidth}{!}{\includegraphics{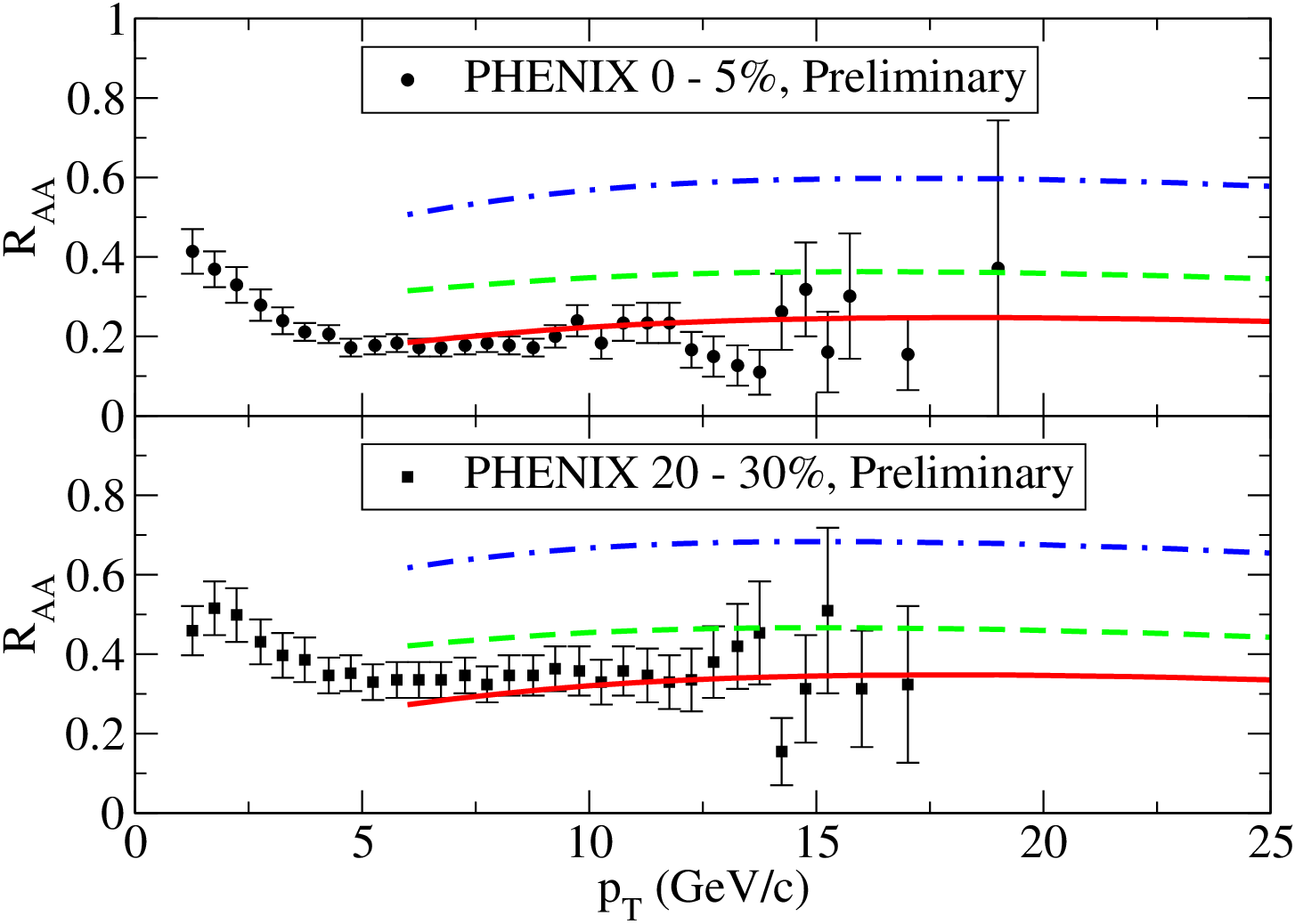}
}\end{center} \caption{(Color online) The nuclear modification factor $R_{AA}$ for neutral pions in central and mid-central collisions (from Ref. \cite{Qin:2007rn}). The dashed curves account for only induced gluon radiation, the dash-dotted curves for only elastic collisions and the solid curves incorporate both elastic and inelastic energy losses.
 } \label{raa}
\end{figure}

In Fig.\ \ref{raa} we show our earlier study \cite{Qin:2007rn} of the
 neutral pion $R_{AA}$ in Au+Au collisions at RHIC measured at
 mid-rapidity for two different impact parameters, $2.4$ fm and $7.5$
 fm, compared with the experimental measurements by PHENIX
 \cite{Adler:2002xw} for the most central (0--5\%) and mid-central
 (20--30\%) collisions. In the calculation, the strong coupling constant
 $\alpha_s$ is adjusted in such a way that the experimental measurement
 of $R_{AA}$ in most central collisions (upper panel) is described. The
 same value, $\alpha_s=0.27$, is used throughout the calculation. (There
 is no additional parameter for the later calculation of high-$p_T$
 photon production and photon-hadron correlations.) Fig.\ \ref{raa} also compares the relative contributions of induced gluon radiations and elastic collisions to the final $R_{AA}$. One may find that while the shape does not show a strong sensitivity, the overall magnitude of $R_{AA}$ is sensitive to both radiative and collisional energy loss.

\begin{figure}[htb]
\begin{center} 
\includegraphics[width=1.0\linewidth]{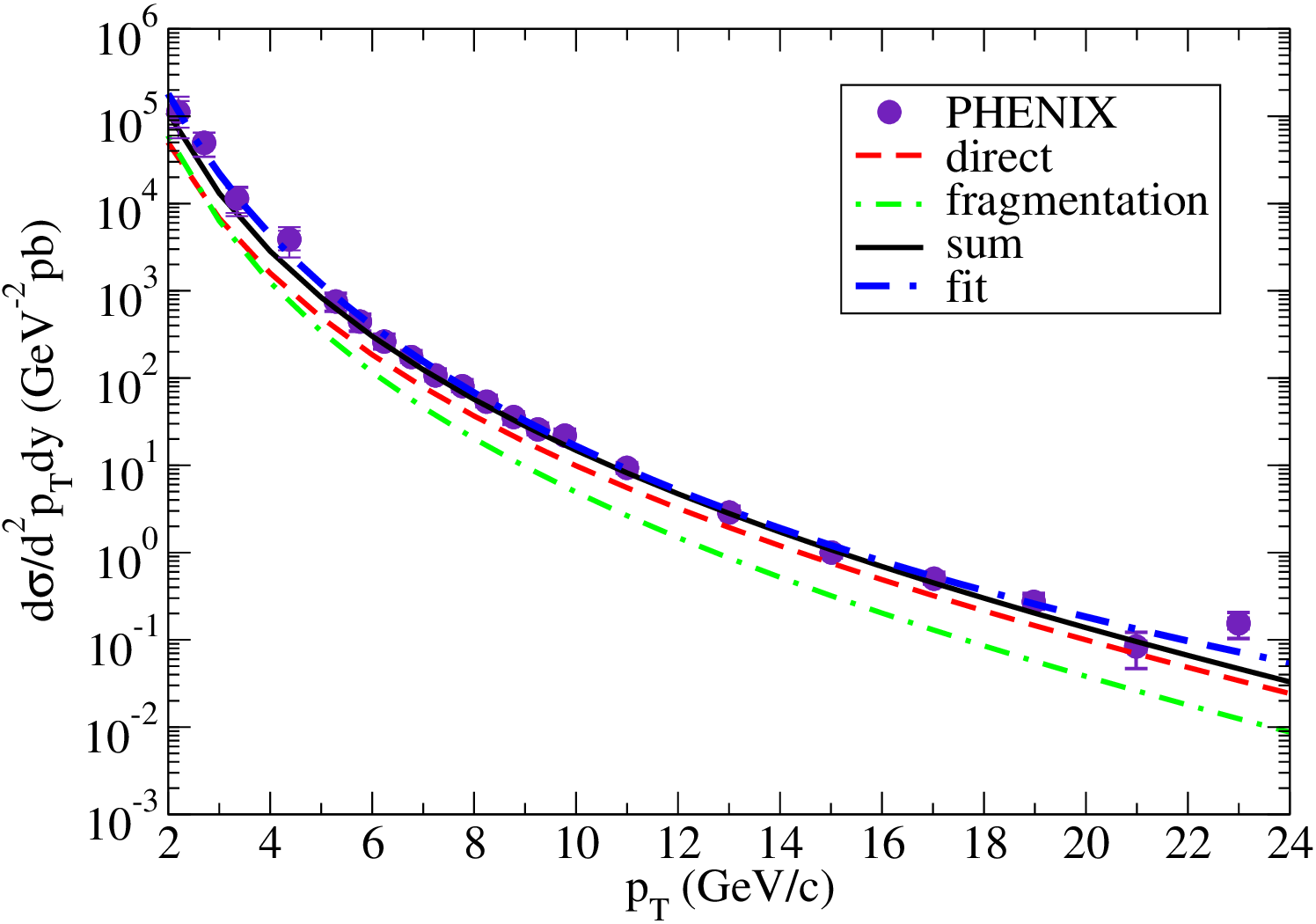}
\end{center}
\caption{(Color online) Photon production in p+p collisions compared with PHENIX data \cite{Adler:2006yt, Adare:2008fq}.} \label{photon_pp}
\end{figure}

Next, we present the results for high-$p_T$ photon production at
RHIC. In Fig.\ \ref{photon_pp}, the inclusive photon cross section in
p+p collisions at $\sqrt{s_{NN}}=200~{\rm GeV}$ as a function of photon
$p_T$ is compared with the experimental measurements by PHENIX
\cite{Adler:2006yt, Adare:2008fq}. The theoretical calculation can
nicely describe the experimental data; this serves as the baseline for
calculating photon production in Au+Au collisions. Photons from early
hard collisions (Compton scatterings and annihilation processes)
dominate at high-$p_T$ regimes, while fragmentation photons gain
increasing significance as photon $p_T$ decreases. Also shown is a
power-law fit to the experimental measurement of total photon yield in
p+p collisions at RHIC: $d\sigma/d^2p_Tdy = 1.027 / (1 + p_T/0.793 \,{\rm GeV})^{6.873} {\rm GeV}^{-2} {\rm mb}$.  This function is somewhat closer to the low $p_{T}$ experimental data, and its usefulness will become clear shortly. 

\begin{figure}[htb]
\begin{center}
\includegraphics[width=1.0\linewidth]{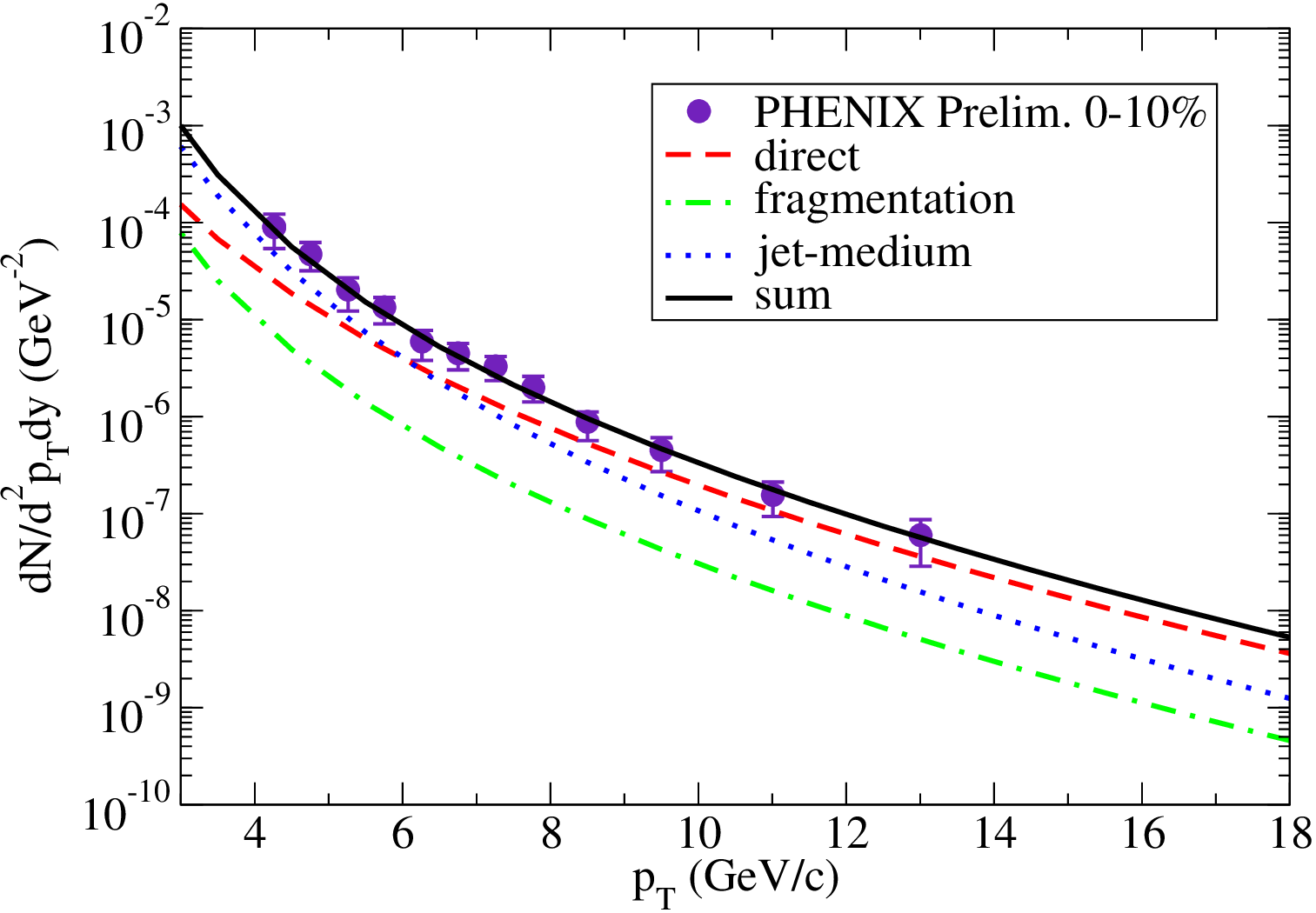}
\end{center}
\caption{(Color online) The contributions from different channels to the photon production in Au+Au collisions at RHIC for $b=2.4$~fm compared with PHENIX data \cite{Isobe:2007ku}. }
\label{photon_AA}
\end{figure}

\begin{figure}[htb]
\begin{center}
\includegraphics[width=1.0\linewidth]{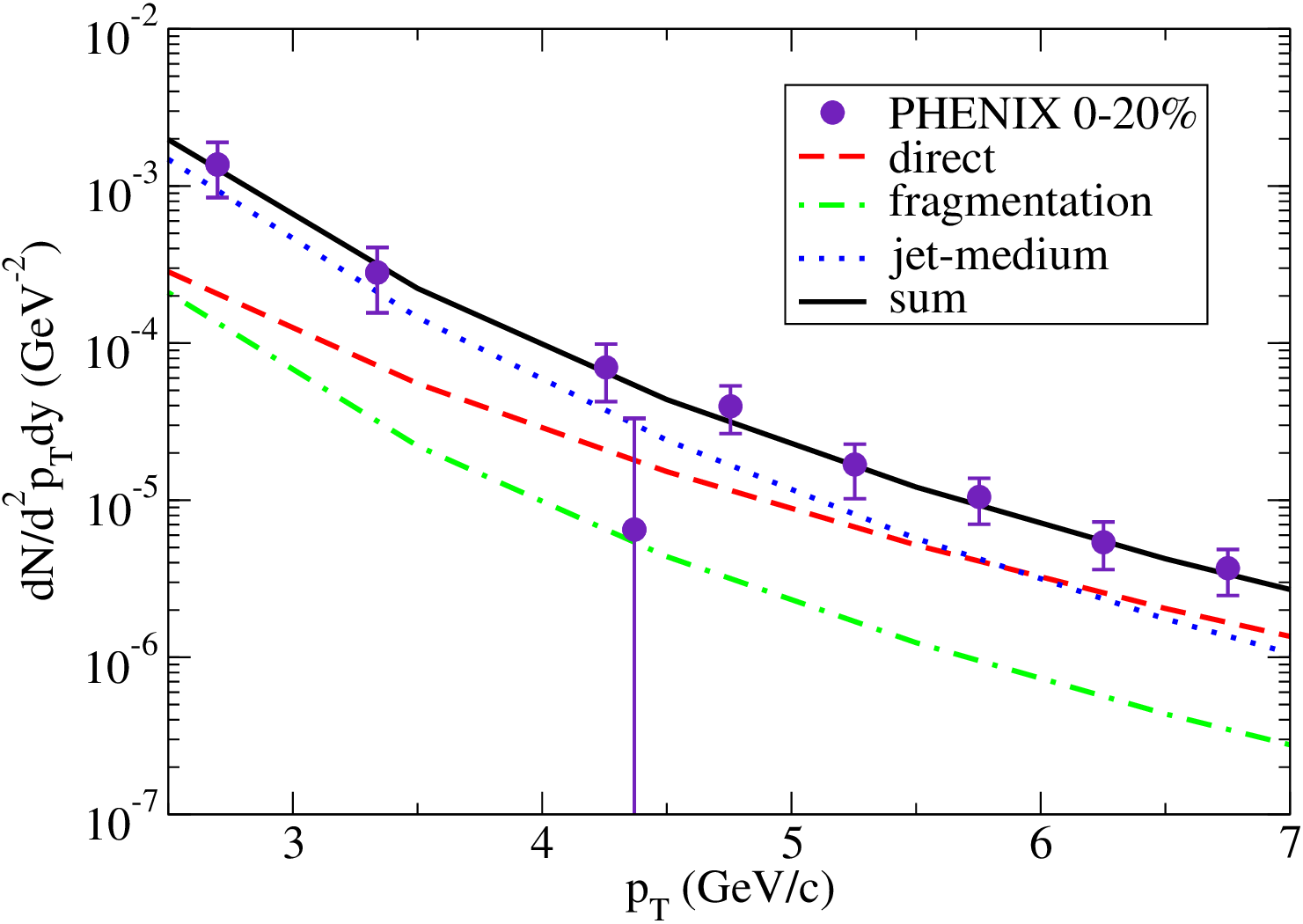}
\end{center}
\caption{(Color online) The contributions from different channels to the photon production in Au+Au collisions at RHIC for $b=4.5$~fm compared with date from the PHENIX collaboration \cite{Isobe:2007ku, Adare:2008fq}.} \label{photon_AA_0-20}
\end{figure}

\begin{figure}[htb]
\begin{center}                                             
\includegraphics[width=1.0\linewidth]{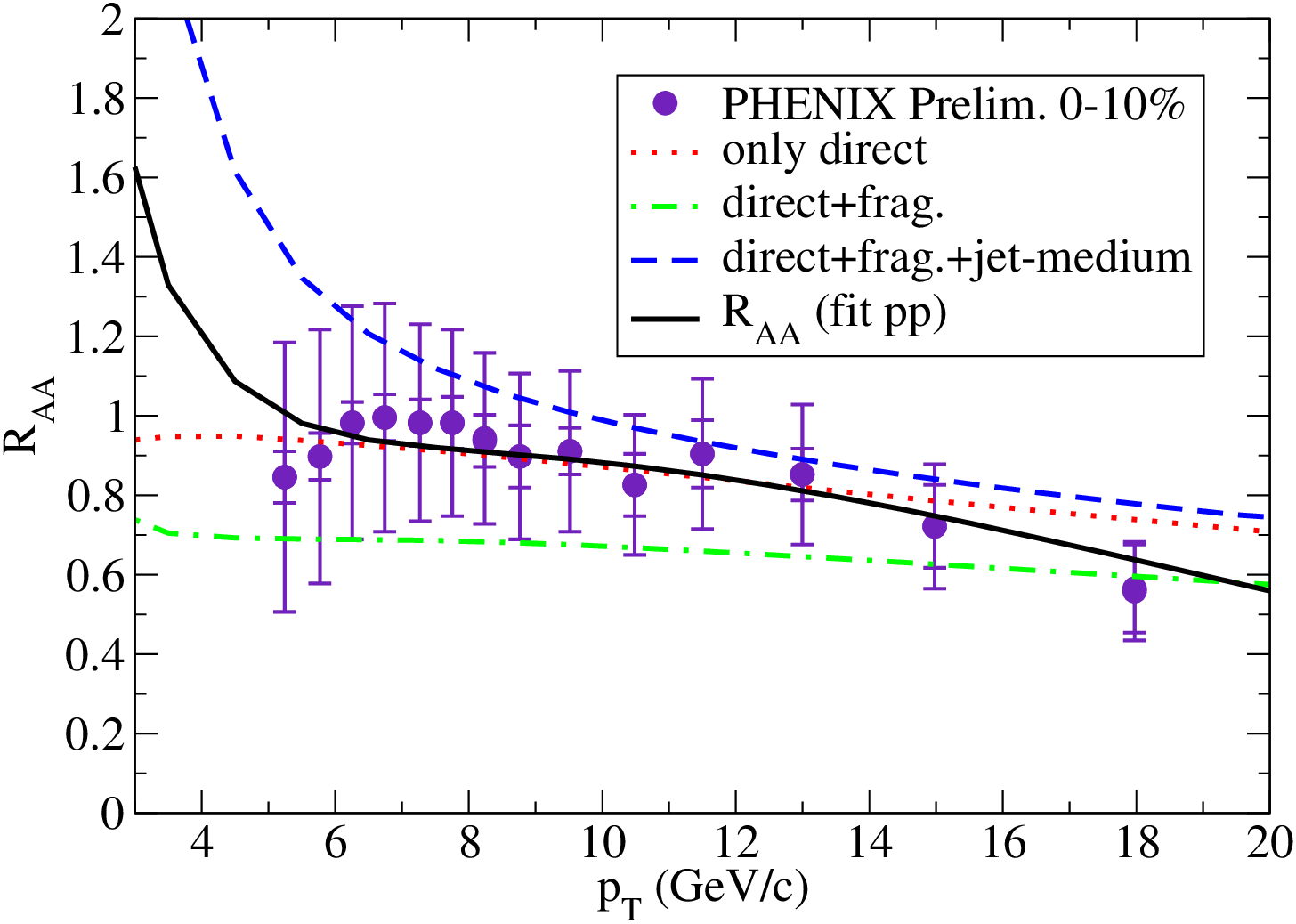}
\end{center}
\caption{(Color online) The nuclear modification factor $R_{AA}$ calculated for photons in Au+Au collisions at RHIC for $b=2.4$~fm compared with PHENIX data with a 0-10\% centrality cut  \cite{Isobe:2007ku}. } \label{photon_raa}
\end{figure}

In Fig.\ \ref{photon_AA}, the relative contributions from different
channels to high-$p_T$ photon production in central Au+Au collisions
($b=2.4$~fm) at RHIC is shown as a function of photon $p_T$ and compared
with most central 0-10\% PHENIX data \cite{Isobe:2007ku}. While direct
photons dominate the high-$p_T$ regime ($p_T \geq 6$~GeV), the presence
of jet-medium interaction is a significant contribution to the total
photon yield in Au+Au collisions at RHIC, especially in the
intermediate-$p_T$ regime ($p_T \approx 3$--$5$~GeV). This can be more
easily seen in Fig.\ \ref{photon_AA_0-20}, where we compare our
calculation of photon production for $b=4.5$~fm in the intermediate
$p_T$ regime with 0-20\% Au+Au collisions data from PHENIX
\cite{Isobe:2007ku, Adare:2008fq}. At very low $p_T$ ($p_T \leq 2$~GeV),
the thermal emissions from partonic and hadronic phases are expected to
dominate \cite{Turbide:2003si}. This component is excluded in the
calculation as we are focusing on the high-$p_T$ regime. Also the
assumptions essential for jet-energy loss calculation break down at such
low $p_T$.

In order to further quantify nuclear medium effects on photon production
in Au+Au collisions, we show in Fig.\ \ref{photon_raa} the calculated photon $R_{AA}$ as a function of photon $p_T$ for central Au+Au collisions
($b=2.4$~fm) at RHIC compared with most 0-10\% PHENIX data. 
The figure shows that $R_{AA}$ can be smaller than $1$ even if only
direct and fragmentation photons are included in both p+p and Au+Au
calculations. As direct photons dominate the total yield (and thus
$R_{AA}$) in the high-$p_T$ regime, the decrease of $R_{AA}$ at high $p_T$ is consistent with an isospin effect. If we include the contribution from fragmentation photons in the calculation, photon $R_{AA}$ gets suppressed because these photons are produced from the fragmentation of the surviving jets with less energies due to the energy loss of jets when they are traversing the thermalized medium. However, the presence of jet-medium interaction in Au+Au collisions (but not in p+p collisions) again enhances $R_{AA}$ if we include them in the total photon yield. It is telling that the sharp rise of $R_{AA}$ at low-$p_T$ originates mainly from the fact that our calculation for p+p collisions is below the data points (see Fig.\ \ref{photon_pp}): a better agreement with the data is obtained with the power-law fit to the p+p data used to calculate $R_{AA}$. The additional QGP sources then manifest themselves in the difference between the solid and the dash-dotted curves in Fig.\ \ref{photon_raa}. The sensitivity of $R_{AA}$ to this level of details also makes clear the need for a precise, QCD-based, quantitative understanding of photon data in p+p collisions.

\begin{figure}[htb]
\begin{center}
\includegraphics[width=1.0\linewidth]{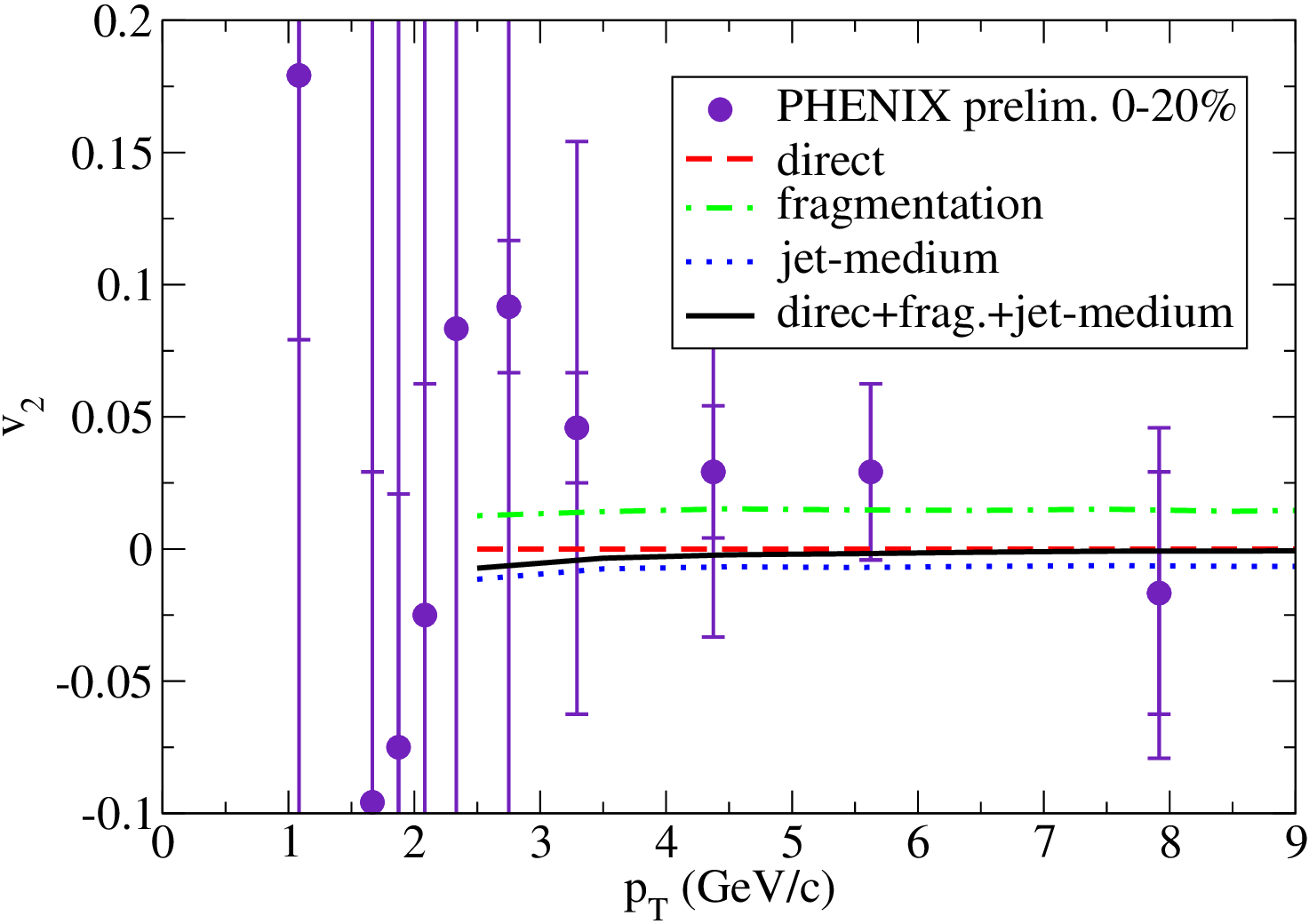}
\end{center}
\caption{(Color online) The photon momentum azimuthal asymmetry, $v_{2}$, as a function of the photon transverse momentum. The different components are explained in the text, and the data are from Ref. \cite{Miki:2008zz}.} \label{photon_v2}
\end{figure}

\begin{figure}[htb]
\begin{center}
\includegraphics[width=1.0\linewidth]{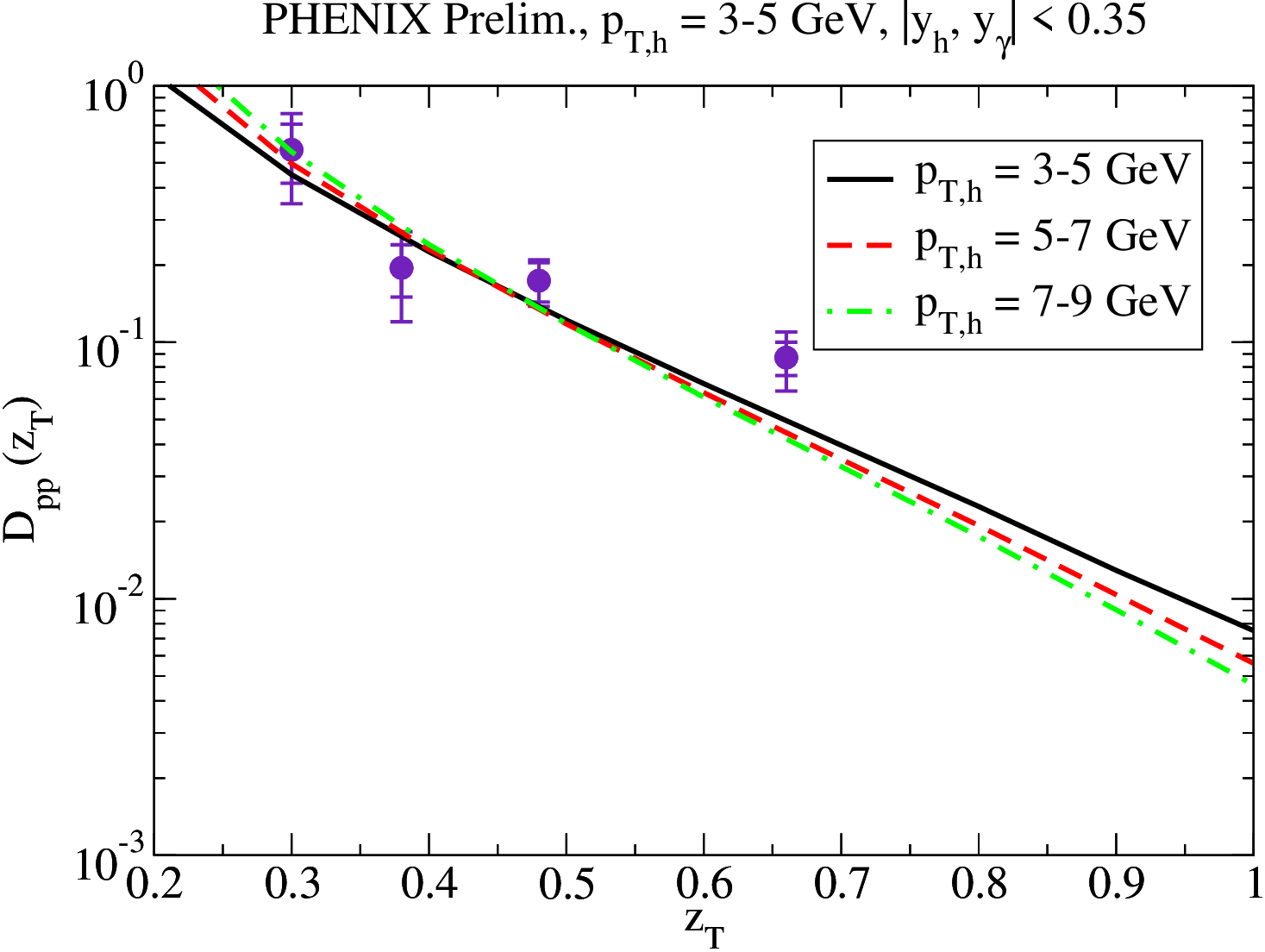}
\end{center}
\caption{(Color online) The photon-triggered fragmentation function as a function of momentum fraction $z_T$ in p+p collisions at RHIC. } \label{phenix_vs_zt}
\end{figure}

\begin{figure}[htb]
\begin{center}
\includegraphics[width=1.0\linewidth]{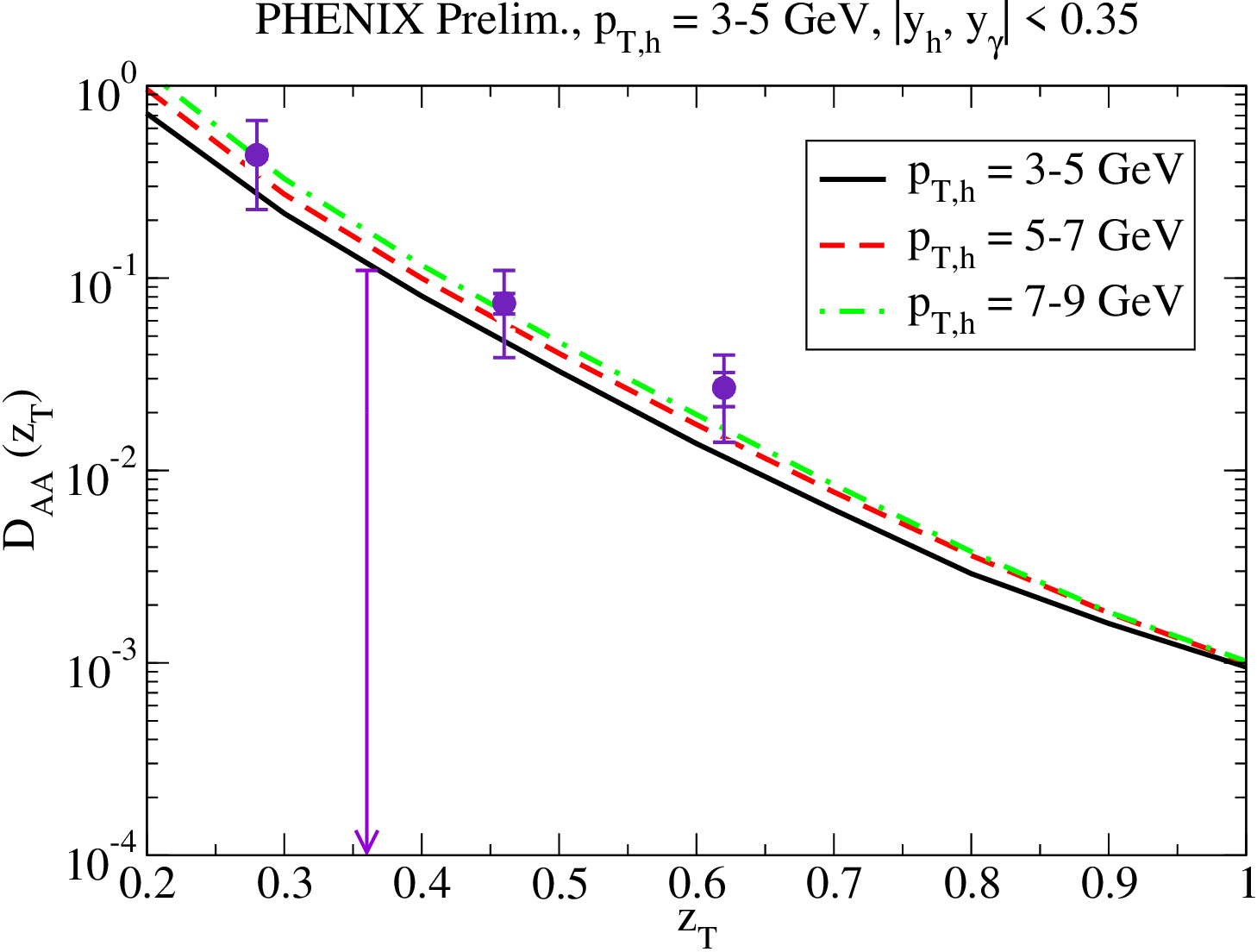}
\end{center}
\caption{(Color online) The photon-triggered fragmentation function as a function of momentum fraction $z_T$ in Au+Au collisions at RHIC, with $b=4.5$~fm. } \label{phenix_vs_zt_aa}
\end{figure}

\begin{figure}[htb]
\begin{center}
\includegraphics[width=1.0\linewidth]{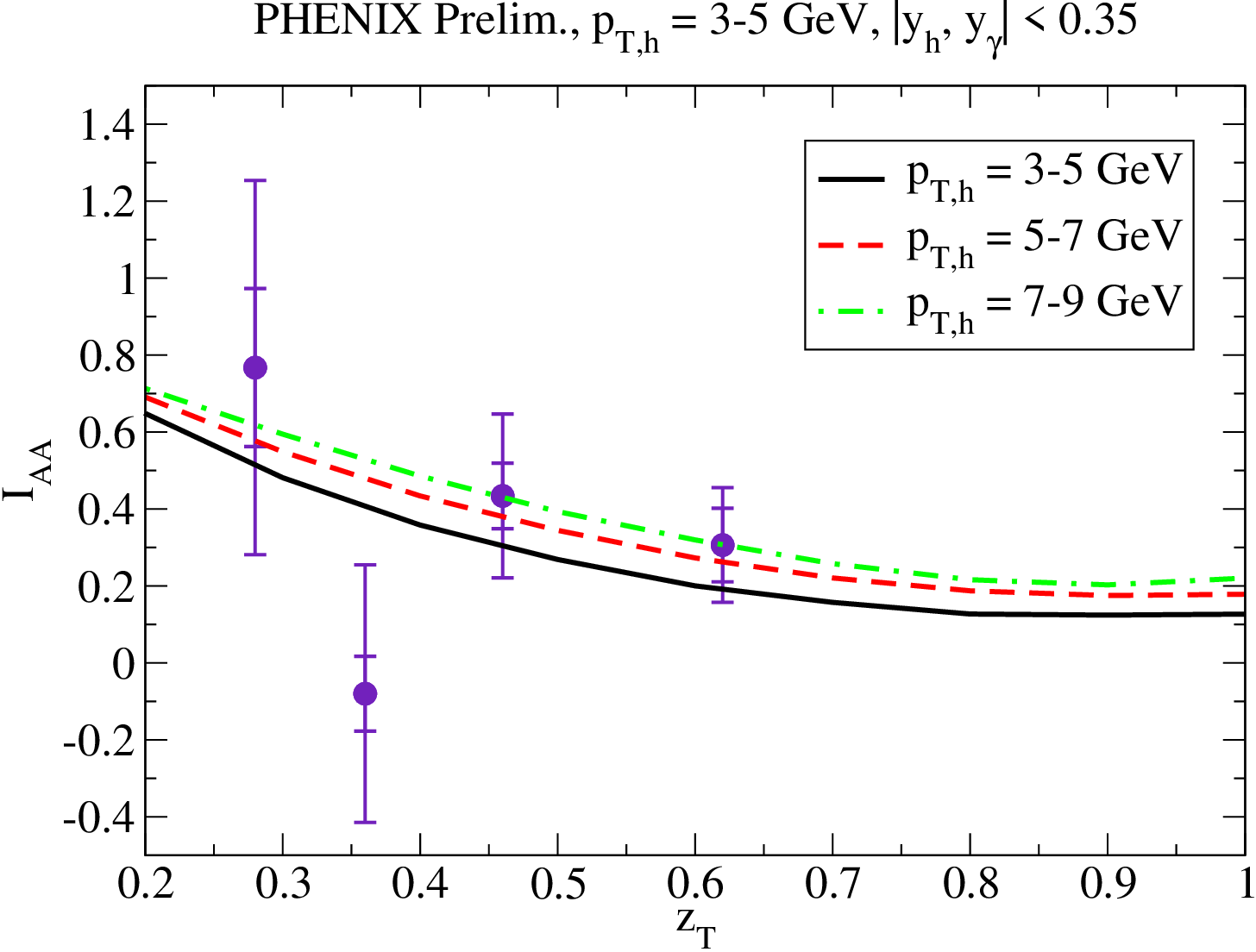}
\end{center}
\caption{(Color online) Photon-triggered $I_{AA}$ as a function of momentum fraction $z_T$ in Au+Au collisions at RHIC, with  $b=4.5$~fm. } \label{phenix_vs_zt_iaa}
\end{figure}

The nucleus-nucleus photon data can also be analyzed in terms of its anisotropy in momentum space, which is characterized by the well-known $v_{2}$ coefficient. This data is show in Figure \ref{photon_v2}. It is clear that the high centrality cut will cause the momentum anisotropy to be small, and this is indeed what is observed. Moreover, the intrinsically negative $v_{2}$ associated with jet-medium photons is drastically reduced in absolute magnitude when combined with the other components whose momentum anisotropy is either zero or positive. Recall that the thermal photons, not included in this calculation, will kick in as the transverse momentum gets lower than $p_{T} \sim 3 - 4$ GeV/c, and that their $v_{2}$ is positive \cite{Chatterjee:2005de,Turbide:2007mi}. A combination of isolation and centrality cuts will be needed to clearly isolate the predicted \cite{Turbide:2005bz} negative contribution \cite{Turbide:2007mi}.

So far, we have just considered single jets and single photons. Now,  results for the correlation between back-to-back high-$p_T$ trigger photons and associated hadrons will be discussed.
In Fig.\ \ref{phenix_vs_zt}, the photon-triggered fragmentation function $D_{pp}(z_T)$  in p+p collisions at RHIC is plotted as a function of momentum fraction $z_T$ for three different associated hadron $p_T$ ranges: $p_T^h = 3$--$5$~GeV, $5$--$7$~GeV and $7$--$9$~GeV. The experimental data are taken from PHENIX \cite{Collaboration:2009vd}, and we have chosen those data points with highest transverse momentum for associated hadrons ($p_T^h = 3$--$5$~GeV) such that fragmentation might be dominant for hadron production. We find that the slopes of the photon-triggered fragmentation function $D_{pp}(z_T)$ are slightly different for the three hadron $p_T$ ranges. We may trace this difference back to the different momentum (fraction) dependence of initial parton distribution functions (PDF) for quarks and gluons (a steeper slope for gluon PDF than quark PDF as a function of momentum fraction).
As we increase the momenta of initial partons (and therefore Bjorken $x$)
more quarks appear in the initial state, so the relative importance of $q\bar{q} \rightarrow
g\gamma$ increases compared to $qg \rightarrow q\gamma$.
As a consequence, the photon-triggered fragmentation function
$D_{pp}(z_T)$, which is an average over quark and gluon fragmentation
functions weighted with their fractions, will become steeper, since the
gluon fragmentation function is steeper than that for quarks.

In Fig.\ \ref{phenix_vs_zt_aa}, we show our results for the
photon-triggered fragmentation function $D_{AA}(z_T)$ in mid-central
Au+Au collisions ($b=4.5$~fm) as a function of momentum fraction $z_T$,
compared with 0-20\% data from PHENIX \cite{Collaboration:2009vd}. The
dependence of $D_{AA}(z_T)$ on associated hadron $p_T$ is similar to
$D_{pp}(z_T)$ in the low $z_T$ regime because the associated hadrons are mostly produced from those jets triggered by direct photons (see Fig.\ \ref{contribution_8-16}). However, this difference tends to decrease at high $z_T$ due to the presence of jet-medium interaction in Au+Au collisions. Since jet-plasma photons and fragmentation photons assume more significance to photon production at lower $p_T$ (see Fig.\ \ref{photon_AA_0-20}), a larger enhancement is obtained for smaller associated hadron $p_T$.

To further quantify the nuclear medium effect on the photon-triggered
hadron production, we may take the ratio of $D_{AA}$ to $D_{pp}$ and
calculate the nuclear modification factor $I_{AA}$ for photon-triggered
hadron production. The result is shown in Fig.\ \ref{phenix_vs_zt_iaa}
and compared with experimental measurement by PHENIX. At low $z_T$,
$I_{AA}$ falls with increasing $z_T$ due to the dominance of direct
photons, but it flattens out at higher $z_T$ due to the effect of
jet-medium photons and fragmentation photons (see Fig.\
\ref{contribution_8-16}).

\begin{figure}[htb]
\begin{center}
\includegraphics[width=1.0\linewidth]{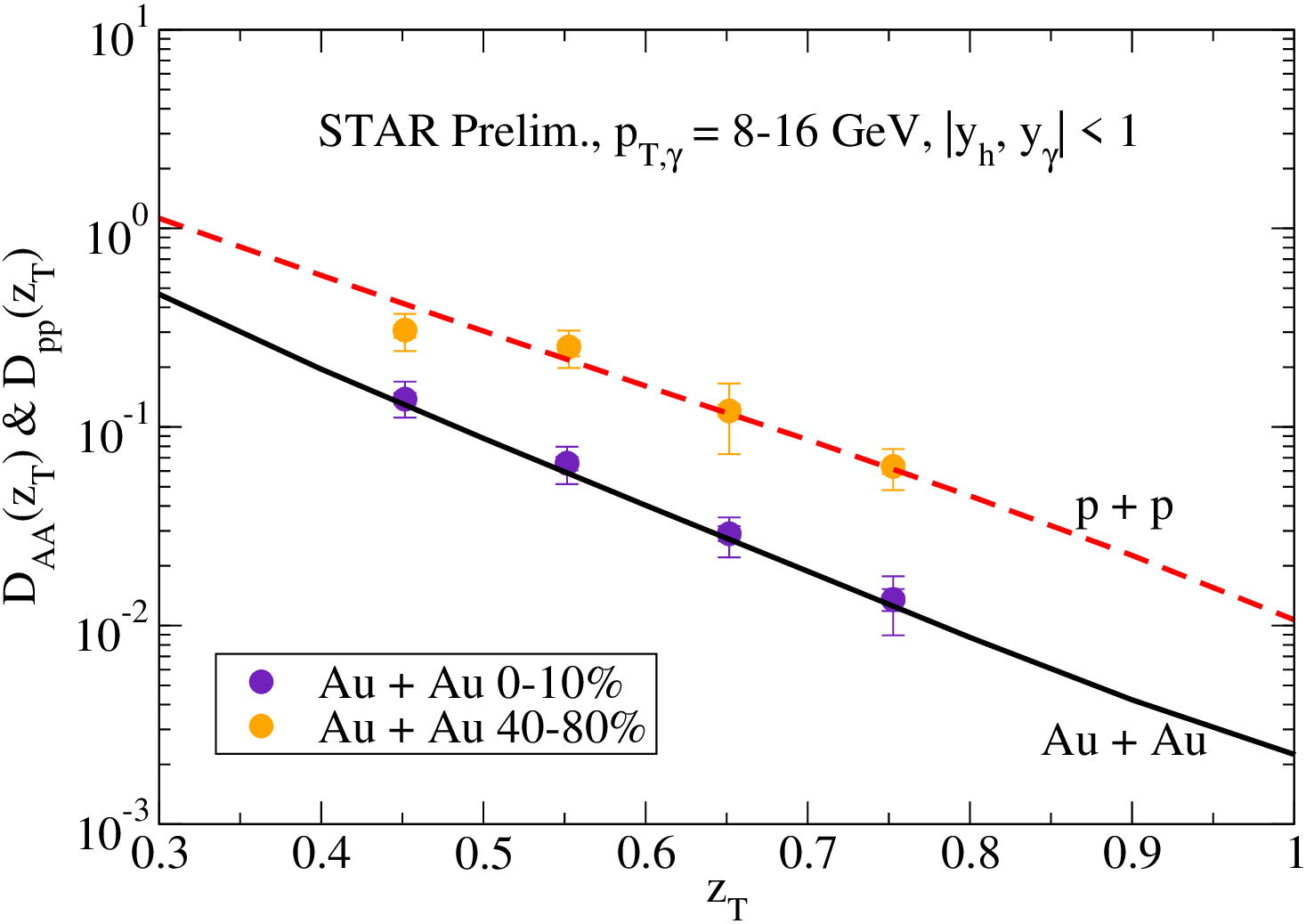}
\end{center}
\caption{(Color online) The photon-triggered fragmentation functions as a function of momentum fraction $z_T$ in central
Au+Au collisions and p+p collisions (peripheral Au+Au collisions) at RHIC. } \label{star_vs_zt}
\end{figure}

\begin{figure}[htb]
\begin{center}
\includegraphics[width=1.0\linewidth]{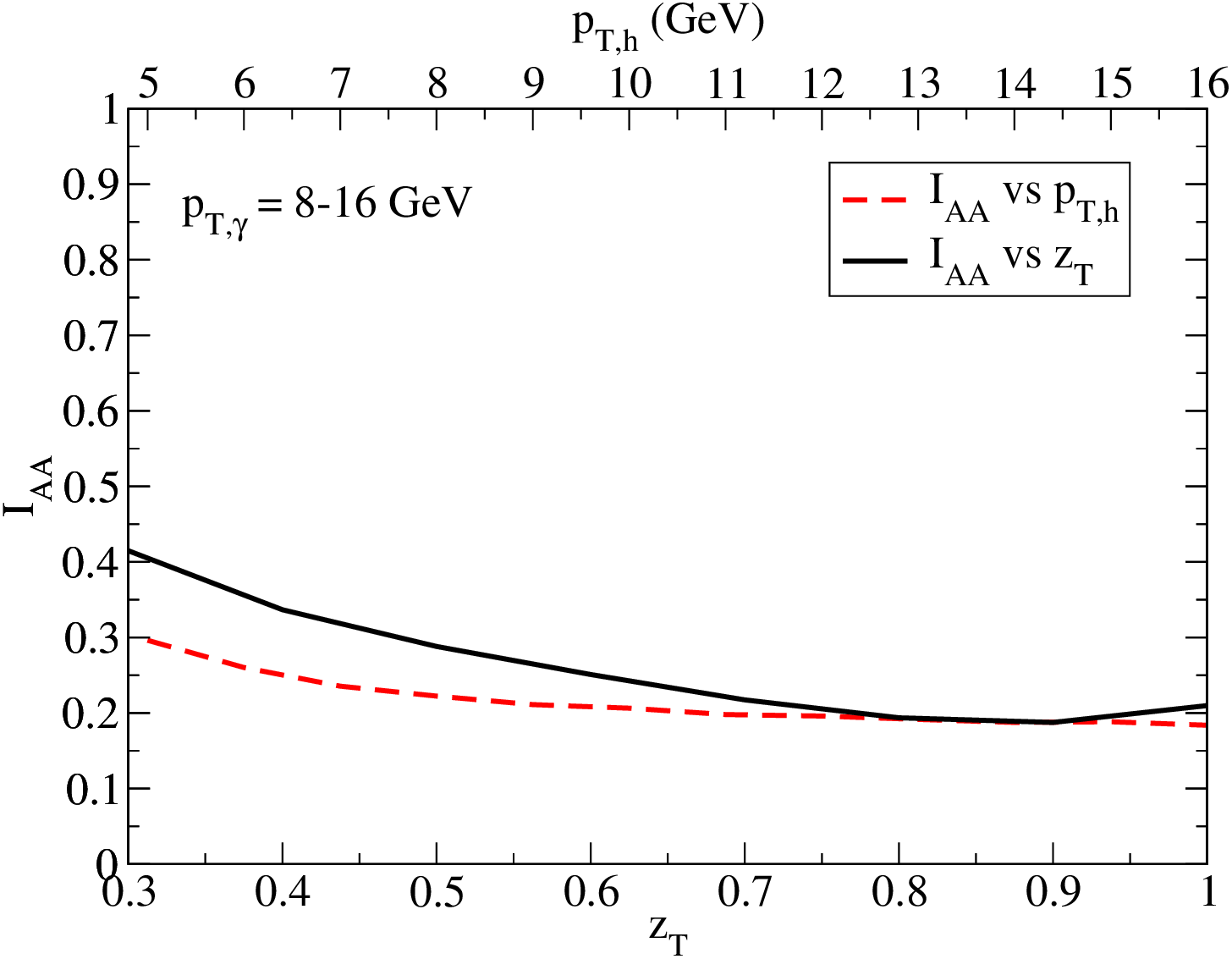}
\end{center}
\caption{(Color online) Photon-triggered $I_{AA}$ for hadrons as a function of hadron momentum $p_T$ or momentum fraction $z_T$ in
Au+Au collisions at RHIC, with $b=2.4$~fm. } \label{iaa_8-16}
\end{figure}

\begin{figure}[htb]
\begin{center}
\includegraphics[width=1.0\linewidth]{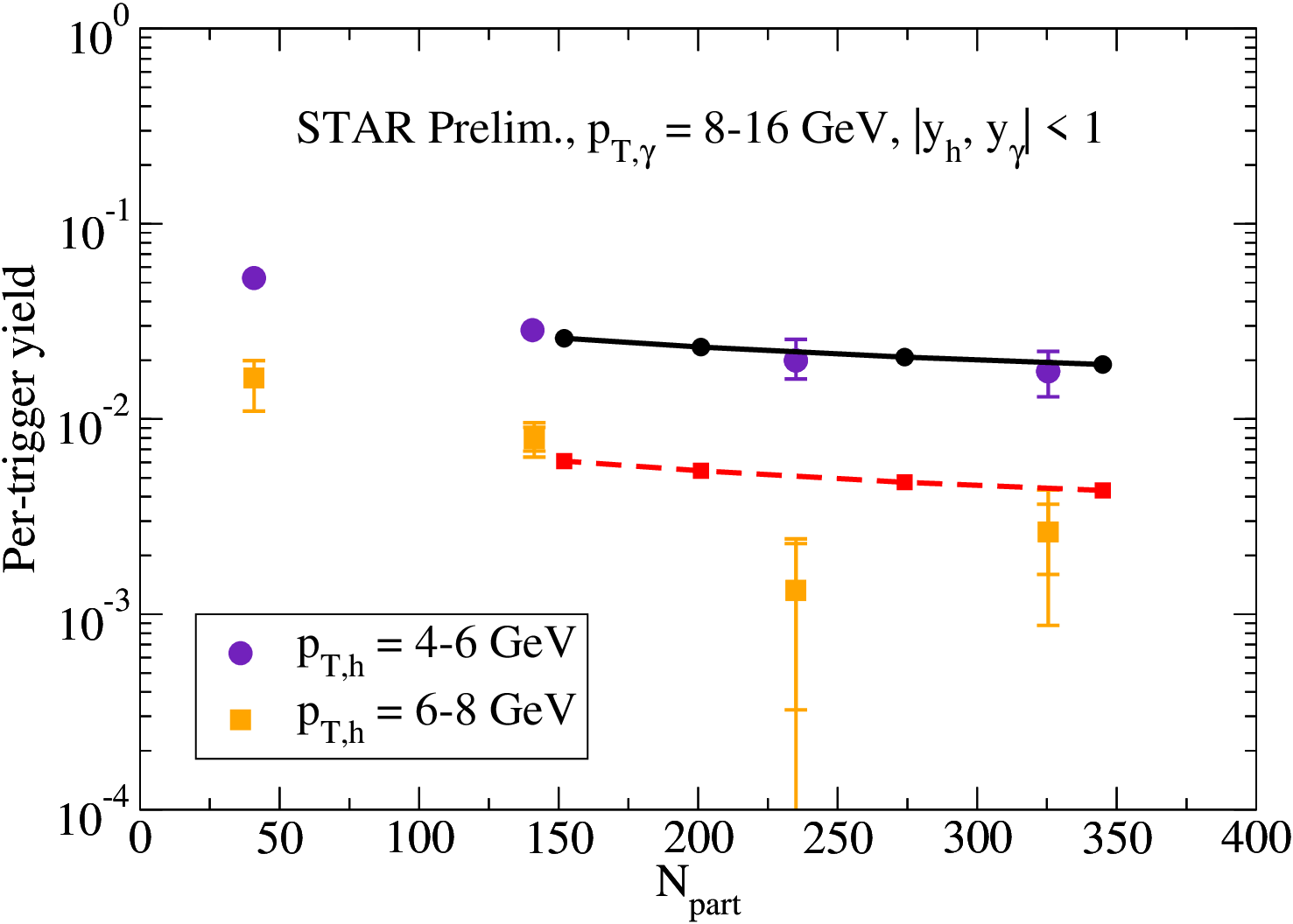}
\end{center}
\caption{(Color online) The per-trigger yield for photon-tagged hadrons in Au+Au collisions at RHIC as a function of
centrality. The four points in each theoretical curve correspond to four impact parameters, $2.4$~fm , $4.5$~fm,
$6.3$~fm and $7.5$~fm. } \label{star_yield}
\end{figure}

Also, we compare our calculation with STAR measurements
\cite{Hamed:2008yz}. In Fig.\ \ref{star_vs_zt}, we show the
photon-triggered fragmentation function $D_{AA}(z_T)$ as a function of
momentum fraction $z_T$ in central Au+Au collisions ($b=2.4$~GeV) at
RHIC. The trigger photon momenta have been chosen to be $p_T^\gamma =
8$--$16$~GeV. The theoretical curve of photon-triggered fragmentation
function $D_{AA}(z_T)$ in central Au+Au collisions agrees with
experimental data quite well. The data for peripheral (40-80\%) Au+Au
collisions are also shown in the figure for the interest of comparing
with the p+p calculation. 
For completeness, the nuclear modification factor $I_{AA}$ for photon-triggered hadron production is plotted in Fig.\ \ref{iaa_8-16} as a function of momentum fraction $z_T$ and hadron momentum $p_T^h$. While $I_{AA}$ as a function of $z_T$ is similar to Fig.\ \ref{phenix_vs_zt_iaa}, $I_{AA}$ is a rather flat function of $p_T^h$ because fixing $p_T^h$ with a wide range of $p_T^\gamma$ is actually an average over a variety of $z_T$.

We also study the medium-size dependence of photon-hadron correlations
by calculating the photon-triggered hadron production in non-central
collisions. By integrating over the transverse momenta of both trigger
photons and associated hadrons, we study the centrality dependence of
the per-trigger yield of photon-triggered hadrons in Au+Au collisions at
RHIC. This result is plotted in Fig.\ \ref{star_yield}, where the
experimental measurements are from STAR \cite{Hamed:2008yz}. In the calculation, we have chosen four different impact parameters, $2.4$~fm, $4.5$~fm, $6.3$~fm and $7.5$~fm, which correspond to four points in each theoretical curve. We do not perform the calculation for very peripheral collisions because the assumption of a thermalized medium essential for a hydrodynamical treatment is no longer fulfilled. Again, the trigger photon $p_T$ has been chosen to be $p_T^\gamma=8$--$16$~GeV, and two different hadron $p_T$ ranges  ($p_T^h = 4$--$6$~GeV and $6$--$8$~GeV) are studied. We find that the centrality dependence of per-trigger yield from the calculation is also consistent with current experimental measurements.

\begin{figure}[htb]
\begin{center}
\includegraphics[width=1.0\linewidth]{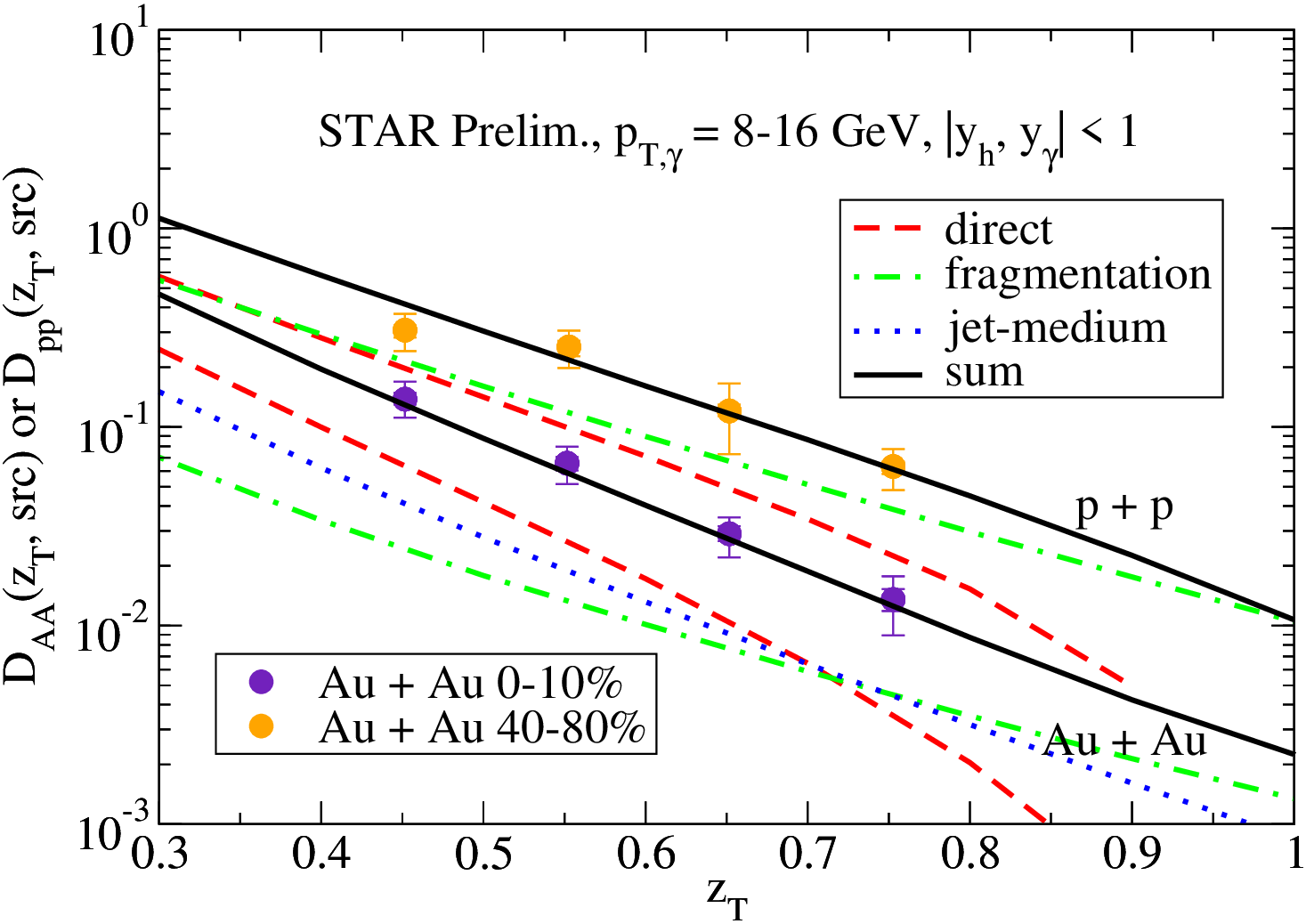}
\end{center}
\caption{(Color online) The contributions from different photon channels to the photon-triggered fragmentation function in  p+p collisions and Au+Au collisions at RHIC.}
\label{contribution_8-16}
\end{figure}

In the above results, we have taken into account all possible sources of high-$p_T$ photon production. It would be interesting to study how different sources of photons contribute the final photon-triggered hadron production.
To address this issue, we may decompose the per-trigger yield of hadrons into different parts, each associated with a specific photon source,
\begin{eqnarray}
P(p_T^h|p_T^\gamma) = \sum_{\rm src} P(p_T^h, {\rm src}|p_T^\gamma),
\end{eqnarray}
where $P(p_T^h, {\rm src}|p_T^\gamma) = P(p_T^h, p_T^\gamma, {\rm src}) / P(p_T^\gamma)$.
A similar decomposition may be performed for photon-triggered fragmentation functions.
In Fig.\ \ref{contribution_8-16}, we show the relative contributions from different photon sources to the photon-triggered fragmentation function $D(z_T, p_T^\gamma)$) in both p+p collisions and central Au+Au collisions ($b=2.4$~fm) if we trigger on a photon on the near side. We find that about half of the away-side hadrons at relatively small $z_T$ are produced from those jets tagged by direct photons, while at higher $z_T$, a large amount of away-side hadrons come from those jets tagged by jet-medium photons and fragmentation photons. Especially, close to $z_T = 1$, where the associated hadrons have almost the same amount of transverse momentum as the trigger photon, the away side hadron production is dominated by those jets tagged by fragmentation photons.
Therefore, it would be interesting to extend the experimental data to
higher $z_T$ (and to high $p_T^h$ such that fragmentation dominates
hadron production), where jet-medium interaction and fragmentation make a significant contribution to photon-hadron correlations.

\section{Conclusions}

In this work, we have studied jet energy loss and photon production at high $p_T$ together in the same framework, with additional information provided by photon-hadron correlations. The energy loss of hard jets traversing the hot and dense medium is computed by a consistent incorporation of both induced gluon radiation and elastic collisions. The production of high $p_T$ photons is obtained by taking into account a complete set of photon-production channels. Numerical results have been presented and compared with experimental measurements by employing a fully (3+1)-dimensional hydrodynamical evolution model for the description of the thermalized medium created at RHIC.

Our results illustrate that the magnitude of jet quenching in relativistic nuclear collisions is sensitive to the inclusion of both radiative and collisional energy loss. It is also found that the interaction between jets and the thermalized medium makes a significant contribution to the total photon production at RHIC. Therefore, it is important to incorporate all sources of hard photons for a full understanding of the correlations between back-to-back hard photons and hadrons. In summary, our study provides the groundwork to experimentally test our understanding of the jet-medium interaction and further extract detailed information about the hot and dense medium created at RHIC.

\section{Acknowledgments}

We are indebted to C. Nonaka and S. Bass for providing their hydrodynamical evolution calculation
\cite{Nonaka:2006yn}. We thank T. Renk and M. Tannenbaum for interesting discussions, and A. M. Hamed and B. Schenke for interesting discussions and a critical reading of this manuscript. This work was supported in part by the U.S. Department of Energy under grant DE-FG02-01ER41190, and in part by the
Natural Sciences and Engineering Research Council of Canada.



\bibliography{qin_reference_list}
\bibliographystyle{h-physrev5}
\end{document}